\documentclass[reprint,amssymb,amsmath,aip,apl]{revtex4-1}
\usepackage{newtxtext,newtxmath}
\usepackage[utf8]{inputenx}
\usepackage{graphicx}
\usepackage{booktabs}
\usepackage{mathrsfs}
\usepackage[dvipsnames]{xcolor}
\usepackage{chemist}
\usepackage[%pdftex,
unicode=true,bookmarks=false,breaklinks=false,pdfborder={0 0 1},colorlinks=true]{hyperref}
\hypersetup{linkcolor=blue,citecolor=blue,urlcolor=blue}
\usepackage[textwidth=17.5cm,textheight=23.5cm,verbose,pdftex]{geometry}
\usepackage{soul}

\newcommand{\fga}{$\phi_{\text{Ga}}$}
\newcommand{\fgas}{$\phi_{\text{Ga$_2$O}}$}

\newcommand{\fo}{$\phi_{\text{O}}$}
\newcommand{\fn}{$\phi_{\text{N}}$}

\newcommand{\tg}{$T_G$}
\newcommand{\alo}{Al$_2$O$_3$}
\newcommand{\gao}{Ga$_2$O$_3$}
\newcommand{\gso}{Ga$_2$O}

\newcommand{\ino}{In$_2$O$_3$}
\newcommand{\sno}{SnO$_2$}
\newcommand{\algao}{(Al$_x$Ga$_{1-x}$)$_2$O$_3$}

\newcommand{\utg}{$\,^{\circ}\text{C}$}

\newcommand{\gr}{\text{growth rate}}

\begin{document}
\graphicspath{{./figs/}}

\title{Adsorption-Controlled Growth of \gao\ by Suboxide Molecular-Beam Epitaxy}
\author{Patrick Vogt}
\email[Electronic mail: ]{pv269@cornell.edu}
\affiliation{Department of Materials Science and Engineering, Cornell University, Ithaca, New York 14853, USA}
\author{Felix V.~E.~Hensling}
\affiliation{Department of Materials Science and Engineering, Cornell University, Ithaca, New York 14853, USA}
\author{Kathy Azizie}
\affiliation{Department of Materials Science and Engineering, Cornell University, Ithaca, New York 14853, USA}
\author{Celesta S.~Chang}
\affiliation{School of Applied and Engineering Physics, Cornell University, Ithaca, New York 14853, USA}
\author{David Turner}
\affiliation{Air Force Research Laboratory, Materials and Manufacturing Directorate, Wright Patterson AFB, Ohio 45433, USA}
\affiliation{Azimuth Corporation, 2970 Presidential Drive, Suite 200, Fairborn, Ohio 45324, USA}
\author{Jisung Park}
\affiliation{Department of Materials Science and Engineering, Cornell University, Ithaca, New York 14853, USA}
\author{Jonathan P.~McCandless}
\affiliation{School of Electrical and Computer Engineering, Cornell University, Ithaca, New York 14853, USA}
\author{Hanjong Paik}
\affiliation{Department of Materials Science and Engineering, Cornell University, Ithaca, New York 14853, USA}
\affiliation{Platform for the Accelerated Realization, Analysis and Discovery of Interface Materials (PARADIM), Cornell University, Ithaca, New York 14853, USA}
\author{Brandon J. Bocklund}
\affiliation{Department of Materials Science and Engineering, Pennsylvania State University, University Park, Pennsylvania 16802, USA}
\author{Georg Hoffman}
\affiliation{Paul-Drude-Institut für Festk{\"o}rperelektronik, Leibniz-Institut im Forschungsverbund Berlin e.V., Hausvogteiplatz 5--7,10117 Berlin, Germany}
\author{Oliver Bierwagen}
\affiliation{Paul-Drude-Institut für Festk{\"o}rperelektronik, Leibniz-Institut im Forschungsverbund Berlin e.V., Hausvogteiplatz 5--7,10117 Berlin, Germany}
\author{Debdeep Jena}
\affiliation{Department of Materials Science and Engineering, Cornell University, Ithaca, New York 14853, USA}
\affiliation{School of Electrical and Computer Engineering, Cornell University, Ithaca, New York 14853, USA}
\affiliation{Kavli Institute at Cornell for Nanoscale Science, Ithaca, New York 14853, USA}
\author{Huili G.~Xing}
\affiliation{Department of Materials Science and Engineering, Cornell University, Ithaca, New York 14853, USA}
\affiliation{School of Electrical and Computer Engineering, Cornell University, Ithaca, New York 14853, USA}
\affiliation{Kavli Institute at Cornell for Nanoscale Science, Ithaca, New York 14853, USA}
\author{Shin Mou}
\affiliation{Air Force Research Laboratory, Materials and Manufacturing Directorate, Wright Patterson AFB, Ohio 45433, USA}
\author{David A.~Muller}
\affiliation{School of Applied and Engineering Physics, Cornell University, Ithaca, New York 14853, USA}
\affiliation{Kavli Institute at Cornell for Nanoscale Science, Ithaca, New York 14853, USA}
\author{Shun-Li Shang}
\affiliation{Department of Materials Science and Engineering, Pennsylvania State University, University Park, Pennsylvania 16802, USA}
\author{Zi-Kui Liu}
\affiliation{Department of Materials Science and Engineering, Pennsylvania State University, University Park, Pennsylvania 16802, USA}
\author{Darrell G.~Schlom}
\email[Electronic mail: ]{schlom@cornell.edu}
\affiliation{Department of Materials Science and Engineering, Cornell University, Ithaca, New York 14853, USA}
\affiliation{Kavli Institute at Cornell for Nanoscale Science, Ithaca, New York 14853, USA}
\affiliation{Leibniz-Institut f{\"u}r Kristalls{\"u}chtung, Max-Born-Str.~2, 12489 Berlin, Germany}

\begin{abstract}
This paper introduces a growth method---suboxide molecular-beam epitaxy ($S$-MBE)---which enables the growth of \gao\ and related materials at growth rates exceeding $1\,\upmu\text{m}\,\text{hr}^{-1}$ with excellent crystallinity in an adsorption-controlled regime. Using a Ga + \gao\ mixture with an oxygen mole fraction of $x(\text{O}) = 0.4$ as an MBE source, we overcome kinetic limits that had previously hampered the adsorption-controlled growth of \gao\ by MBE. We present growth rates up to $1.6\,\upmu\text{m}\,\text{hr}^{-1}$ for \gao/\alo\ heterostructures with unprecedented crystalline quality and also at unparalleled low growth temperature for this level of perfection. We combine thermodynamic knowledge of how to create molecular-beams of targeted suboxides with a kinetic model developed for the $S$-MBE of III-VI compounds to identify appropriate growth conditions. Using $S$-MBE we demonstrate the growth of phase-pure, smooth, and high-purity homoepitaxial \gao\ films that are thicker than $4\,\upmu\text{m}$. With the high growth rate of $S$-MBE we anticipate a significant improvement to vertical \gao-based devices. We describe and demonstrate how this growth method can be applied to a wide-range of oxides. $S$-MBE rivals leading synthesis methods currently used for the production of \gao-based devices.
\end{abstract}

%\date{\today}

\maketitle
% Samples used in this work:
% Fig.1
% Fig.2
% Fig.3
\section{Introduction}
\setlength{\parindent}{5ex}
Molecular-beam epitaxy (MBE) involves the growth of epitaxial thin films from molecular-beams.  In `conventional' MBE the molecular-beams consist of elements. An example is the Ga ($g$) species that evaporate from a heated crucible containing Ga ($l$) or the As$_4$ ($g$) species that evaporate from a heated crucible containing As ($s$), where $g$, $l$, and $s$ denote gaseous, liquid, and solid, respectively. In gas-source MBE the species in the molecular-beams originate from gases that are plumbed into the MBE from individual gas cylinders, for example, arsine or phosphine.  In metal-organic MBE the species in the molecular-beams are metal-organic molecules like trimethylgallium or trimethylaluminum.\cite{hermann1996} `Suboxide MBE' refers to an MBE growth process utilizing molecular-beams of suboxides like \gso\ ($g$) or In$_2$O ($g$).  We have applied this method to the growth of \gao\  thin films and find that it can produce epitaxial \gao\ films with far greater perfection and at much higher growth rates than currently demonstrated by other MBE methods for the growth of this material.  

\subsection{`Conventional' MBE of \gao\ and related materials}
\setlength{\parindent}{5ex}
Gallium-sesquioxide (\gao) synthesized in its different polymorphs [i.e., $\upalpha$-\gao\ (rhomboheral), $\upbeta$-\gao\ (monoclinic), $\upgamma$-\gao\ (cubic spinel), $\upepsilon$-\gao\ (hexagonal), and $\upkappa$-\gao\ (orthorhombic)], is an emerging semiconductor possessing promising features for unprecedented high-power electronics. This is due to its large band gap ($\sim 5\,\text{eV}$)\cite{onuma2015,wang2018} and very high breakdown field (up to $8\,\text{MV}\,\text{cm}^{-1}$).\cite{higashiwaki2012} The band gap of \gao\ may be widened by alloying \gao\ with \alo\ to form \algao.\cite{wang2018} The synthesis of \algao/\gao\ heterostructures with high Al content $x$ is desired for high-power transistors with large band gap offsets.\cite{krishna2017,wang2018, jinno2020} 

\setlength{\parindent}{5ex}
It is known that the `conventional' MBE of \gao---i.e., when supplying monoatomic Ga and active O species during growth---is strongly limited by the formation and subsequent desorption of its volatile suboxide \gso.\cite{tsai2010,vogt2015a,vogt2016a,vogt2017a,vogt2018b} In the adsorption-controlled regime (i.e., grown with an excess of Ga), its \gr\ strongly decreases with increasing Ga flux, \fga, because not enough oxygen is available to oxidize the physisorbed \gso\ to \gao\ ($s$) and the \gso\ desorbs from the hot substrate. At sufficiently high \fga, film growth stops, and even goes negative (i.e., the \gao\ film is etched).\cite{vogt2015a} This effect is enhanced as the growth temperature, \tg, increases due to the thermally activated desorption of \gso\ from the growth surface. The enhanced, \tg-induced \gso\ desorption leads to a decreasing \gr\ even in the O-rich regime, resulting in a short \gr\ plateau (the value of which is far below the available active O flux \cite{vogtdiss}), followed by an even further decreasing \gr\ in the adsorption-controlled regime.\cite{vogt2016a,vogt2016b,vogtdiss} These effects, i.e., the O-deficiency induced and thermally activated desorption of suboxides,\cite{vogt2016a,vogt2016b,vogtdiss,vogt2018b} are detrimental for the growth of III-VI (e.g., \gao) and IV-VI materials in the adsorption-controlled regime.

\setlength{\parindent}{6ex}
Nevertheless, the MBE of thin films in the adsorption-controlled growth regime is often desired for high crystal perfection,\cite{migita1997,ulbricht2008,driscoll2020} smooth surface morphology,\cite{bierwagen2009} avoiding undesired oxidation states,\cite{paik2017,mei2019} or suppressing the formation of electrically compensating defects.\cite{korhonen2015,peelaers2019} 

\setlength{\parindent}{6ex}
The decreasing \gr\ of \gao\ is microscopically explained by a complex two-step reaction mechanism.\cite{vogt2018b,vogtdiss} In the \textit{first} reaction step, all Ga oxidizes to \gso\ via the reaction
\begin{equation}
\label{eq:first}
2\text{Ga}\,(a)  + \text{O}\,(a) \; \reactrarrow{0pt}{0.75cm}{}{} \;  \text{Ga$_2$O} \, (a,g)  \;,
\end{equation}
with adsorbate and gaseous phases denoted as $a$ and $g$, respectively.
The \gso\ formed may either desorb from the growth surface (in the O-deficient regime or at elevated \tg) or be further oxidized to \gao\ via a \textit{second} reaction step through the reaction
\begin{equation}
\label{eq:sec}
\text{Ga$_2$O}\,(a)  + 2\text{O}\,(a) \; \reactrarrow{0pt}{0.75cm}{}{} \;  \text{\gao} \, (s)  \;,
\end{equation}
with the solid phase denoted as $s$.

\setlength{\parindent}{6ex}
This two-step reaction mechanism and the resulting \gso\ desorption defines the \gr-limiting step for the `conventional' MBE of \gao\ and related materials.\cite{vogt2018b,vogtdiss} This results in a rather narrow growth window associated with low \gr s in the adsorption-controlled regime.\cite{tsai2010,vogt2015a,vogt2016a,vogt2018b} A similar \gr-limiting behavior, based on this two-step reaction mechanism, has also been reported for the growth of other III-VI (e.g., \ino) and IV-VI (e.g., \sno) compounds by `conventional' MBE.\cite{vogt2015a,vogt2016b,vogt2018b} This two-step growth process for the growth of III-VI and IV-VI oxides by `conventional' MBE is fundamentally different from the single-step reaction mechanism of, for example, III-V \cite{calleja1999,koblmuller2007,garrido2008} and II-VI \cite{kato2003} compounds. It can be attributed to the different electronic configurations of the compound constituents, resulting in different compound stoichiometries between III-VI and IV-VI compared with III-V and II-VI materials, respectively.

\setlength{\parindent}{6ex}
In the growth method introduced in this work, which we call \textit{suboxide} MBE ($S$-MBE), we avoid the first reaction step (\ref{eq:first}) by directly supplying a \gso\ ($g$) molecular-beam to the growth front on the substrate surface. Using this approach, we bypass the \gr-limiting step for `conventional' \gao\ MBE by removing the O-consuming step to \gso\ formation that occurs on the substrate in the `conventional' MBE growth of \gao.\cite{vogtdiss,vogt2018b} A related approach has been used by Ghose \textit{et al}.\cite{ghose2016,ghose2017} with \gso\ provided from \gao\ source material heated to temperatures well in excess of $1600\,^{\circ}\text{C}$ to produce a molecular beam of \gso\ for the growth of \gao\ films by MBE.\cite{droopad2017} Motivated by known vapor pressure data of oxides \cite{lamoreaux1983} and their mixtures with the respective metals, e.g., Ga + \gao,\cite{frosch1962} as well as the possibility of decomposing \gao\ by Ga and \sno\ by Sn under MBE conditions,\cite{vogt2015a} Hoffmann \textit{et al}. \cite{hoffmann2020} have demonstrated how mixtures of Ga with \gao\ and Sn with \sno\ provide MBE-relevant fluxes of \gso\ and SnO, respectively, at source temperatures below $1000\,^{\circ}\text{C}$. This prior work has grown films using suboxide molecular beams by MBE at growth rates $< 0.2 \, \upmu\text{m}\,\text{hr}^{-1}$.\cite{passlack2003,hoffmann2020}

\setlength{\parindent}{6ex}
As we demonstrate, $S$-MBE enables the synthesis of \gao\ in the highly adsorption-controlled regime, at $\text{\gr s} > 1\,\upmu\text{m}\,\text{hr}^{-1}$ with unparalleled crystalline quality for \gao/\alo\ heterostructures as well as homoepitaxial \gao\ at relatively low \tg. The \gr\ of $S$-MBE is competitive with other established growth methods used in semiconductor industry---such as chemical vapor deposition (CVD) \cite{rafique2018} or metal-organic CVD (MOCVD) \cite{zhang2019}---and moreover, leads to better structural perfection of the obtained thin films. With this improved perfection we expect an improvement of $n$-type donor mobilities in \gao\ thin films doped with Sn, Ge, or Si grown by $S$-MBE, as well. The relatively low \tg\ at which it becomes possible to grow high-quality films by $S$-MBE is a crucial enabler for materials integration where temperatures are limited, e.g., back end of line (BEOL) processes.
%%%%%%%%%%%%%%%%%%%%%%%%%%%%%%%%%%%%%%%%%%%%%%%%%%%%%%%%%%%%%%%%%%%%%%%%.
\begin{figure}[t]
\includegraphics[scale=1.7]{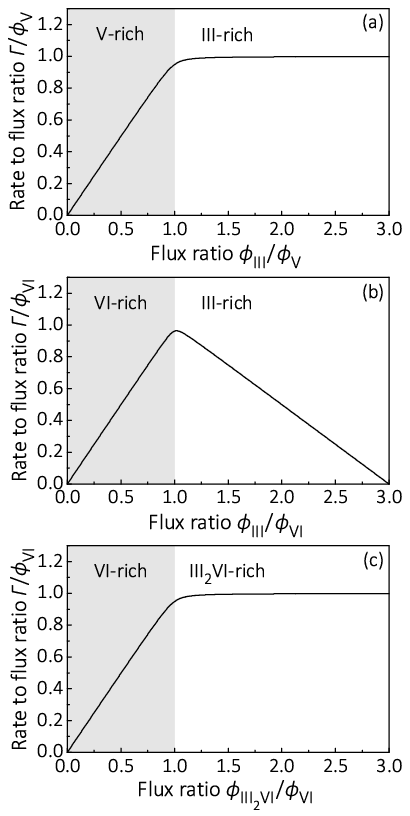}
\caption{(a) and (b) Schematic \gr\ as observed for III-V (e.g., GaN)\cite{garrido2008} and III-VI compounds (e.g., \gao)\cite{vogt2018b} as a function of the III/V (e.g., \fga/\fn) and III/VI flux ratios (e.g., \fga/\fo), respectively. (c) Anticipated \gr\ behavior of III-VI compounds (e.g., \gao)\cite{vogtdiss} as a function of the III$_2$VI/VI flux ratio (e.g., \fgas/\fo). All schematic \gr\ evolutions are normalized by the respective fluxes of active available group V ($\phi_{\text{V}}$) and group VI elements ($\phi_{\text{VI}}$). Each plot is at a constant \tg. Anion-rich and cation-rich regimes are indicated in gray and white, respectively.}
\label{fig:schem}
\end{figure}
%%%%%%%%%%%%%%%%%%%%%%%%%%%%%%%%%%%%%%%%%%%%%%%%%%%%%%%%%%%%%%%%%%%%%%%%

\setlength{\parindent}{6ex}
Figure \ref{fig:schem} illustrates a schematic of how growth rate depends on cation flux during the MBE growth of different types of compounds, where both axes are normalized by the anion flux. Figure \ref{fig:schem}(a) depicts the observed behavior for III-V compounds, e.g., GaN.\cite{garrido2008} Figure \ref{fig:schem}(b) shows the observed behavior for III-VI compounds, e.g., \gao, when the group III cation is supplied by a molecular-beam of the group III element (e.g., Ga).\cite{vogt2015a} In Fig.~\ref{fig:schem}(c), the anticipated behavior for III-VI compounds is plotted, e.g., \gao, when the group III element is supplied by a molecular-beam of a III$_2$VI subcompound containing the group III constituent (e.g., \gso).\cite{vogtdiss} The units of the horizontal and vertical axes are chosen to make the crossover occur at values of unity. For the sake of simplicity, henceforth, we only discuss the reaction behavior of GaN and \gao\ in detail. We emphasize, however, this discussion holds true for the MBE growth of AlN,\cite{calleja1999} InN,\cite{koblmuller2007} \ino\ (Refs.~\citenum{vogt2015a,vogt2016b,vogt2018b}) and other III-VI,\cite{vogt2018a,vogt2018b} and II-VI compounds.\cite{kato2003}

\setlength{\parindent}{6ex}
As drawn in Figs.~\ref{fig:schem}(a)--\ref{fig:schem}(c), the \gr\ of GaN and \gao\ increases linearly with increasing \fga\ in the N-rich [Fig.~\ref{fig:schem}(a)] and O-rich regimes [Fig.~\ref{fig:schem}(b) and \ref{fig:schem}(c)], respectively. Here, the incorporation of Ga is limited by the impinging \fga\ or \gso\ flux, \fgas\ (i.e., Ga-transport and \gso-transport limited growth regimes). 

\setlength{\parindent}{6ex}
For GaN MBE [Fig.~\ref{fig:schem}(a)], once the supplied \fga\ exceeds the flux \fn\ of active available N, the \gr\ saturates, is independent of the \fga/\fn\ ratio, and is limited by \fn\ and \tg. The measured plateau in \gr\ for GaN MBE in the Ga-rich regime results from its single-step reaction kinetics. Here, Ga reacts directly with activated N via the reaction \cite{garrido2008}
\begin{equation}
\label{eq:gan}
\text{Ga}\,(a)  + \text{N}\,(a) \; \reactrarrow{0pt}{0.75cm}{}{} \;  \text{GaN} \, (s)  \;,
\end{equation}
and excess Ga either adsorbs onto or desorbs from the growth surface depending upon \fn\ and \tg.

\setlength{\parindent}{6ex}
Figure \ref{fig:schem}(b) depicts the reaction kinetics of \gao\ in the Ga-rich regime (O-deficient growth regime) by supplying \fga. Here, the \gr\ linearly \textit{decreases} with increasing \fga, and the growth eventually stops at $\text{\fga} \geq 3\text{\fo}$ (in growth rate units). The fact that desorbing \gso\ removes Ga \textit{and} O from the growth surface---that cannot contribute to \gao\ formation---leads to the decreasing \gr\ in the O-deficient growth regime.\cite{vogt2015a,vogt2016a,vogt2018b} This behavior is microscopically governed by the two-step reaction process, Eqs.~(\ref{eq:first}) and (\ref{eq:sec}),\cite{vogt2018b} and is fundamentally different from the single-step reaction kinetics, Eq.~(\ref{eq:gan}), governing the MBE of GaN [Fig.~\ref{fig:schem}(a)].

\setlength{\parindent}{6ex}
In Fig.~\ref{fig:schem}(c), the anticipated growth kinetics of \gao\ while using a \gso\ beam is depicted, showing a constant \gr\ in the \gso-rich regime (i.e., in an excess of \gso).\cite{vogtdiss}  Excess \gso\ (that cannot be oxidized to \gao) either accumulates or desorbs off the growth surface \textit{without} consuming or removing active O from its adsorbate reservoir---similar to the case presented for GaN in Fig.~\ref{fig:schem}(a). Thus, with $S$-MBE, one may effectively achieve single-step reaction kinetics for \gao\ MBE [reaction (\ref{eq:sec})], as is the case for the growth of GaN by MBE [reaction (\ref{eq:gan})].

\setlength{\parindent}{6ex}
The synthesis of III-V and II-VI materials with cation flux-independent growth rates in adsorption-controlled growth regimes---originating from their simple single-step reaction kinetics [e.g., reaction (\ref{eq:gan})]---is beneficial for device-relevant growth rate control and the improvement of their crystal properties.\cite{hey2000,gog2004,neugebauer2003}
Through the use of $S$-MBE, we convert the complex two-step reaction kinetics of III-VI [e.g., reactions (\ref{eq:first}) and (\ref{eq:sec})] and IV-VI compounds into simple single-step kinetics [e.g., (\ref{eq:sec})], the same as observed for III-V and II-VI materials. We therefore expect a similar growth behavior during $S$-MBE, i.e., constant \gr s in the adsorption-controlled regime, which are highly scalable by the provided active O flux. Such a regime should allow III-VI thin films (e.g., \gao) and IV-VI films (e.g., \sno) to be grown much faster with excellent crystal quality at relatively low \tg.

\setlength{\parindent}{6ex}
$S$-MBE utilizes molecular-beams of suboxides and builds upon prior thermodynamic work and thin film growth studies.  For example, molecular-beams of the following suboxides have all been used in MBE:~\gso,\cite{passlack2003,ghose2016,ghose2017} GdO,\cite{kwo2001,bierwagen2013} LuO,\cite{bierwagen2013}, LaO,\cite{bierwagen2013} NdO,\cite{fissel2006} PrO,\cite{liu2001,wata2009} ScO, \cite{chen2005} SnO, \cite{sasaki2012,rag2016,paik2017,mei2019,hoffmann2020} YO. \cite{kwo2001} Even before these MBE studies, thin films of the suboxides SiO, \cite{hass1950,per2009}, SnO, \cite{geurts1984,krase1985,pan2001,pan2001a,guo2010}, and GeO \cite{astan2017} had been deposited by thermal evaporation, exploiting the same underlying vapor pressure characteristics that make $S$-MBE possible.  In some of these cases the dominant species in the gas phase were not identified, but subsequent vapor pressure studies and thermodynamic calculations establish that they were suboxides.\cite{lamoreaux1983,adkison2020}

\setlength{\parindent}{6ex}
What is new about $S$-MBE is the use of suboxide molecular-beams in a targeted way to achieve epitaxial growth of desired oxides (e.g., \gao) at high growth rates in an adsorption-controlled regime.  This enables the benefits of the far simpler (from a growth kinetics, growth control, and growth standpoint) plateau growth regime shown in Fig.~\ref{fig:schem}(c) to be harnessed rather than the growth regime shown in Fig.~\ref{fig:schem}(b) that has posed limits to the growth of \gao\ films by `conventional' MBE up to now. 

\section{Detailed description of $S$-MBE}
\setlength{\parindent}{6ex}
The use of a \gso\ ($g$) molecular-beam to grow \gao\ $(s)$ thin films by MBE in the O-rich regime (i.e., in an excess of active O) has been demonstrated by placing the stoichiometric solid of the compound \gao\ into a crucible and using it as an MBE source.\cite{ghose2016, ghose2017} Possible reactions that produce a \gso\ molecular-beam by the thermal decomposition of \gao\ are:
\begin{align}
\text{\gao}\,(s)  &\; \reactrarrow{0pt}{0.75cm}{}{} \;  \text{\gso} \, (a,g) +  \text{O}_2 \, (a,g) \label{eq:c1} \\
\text{\gao}\,(s)  &\; \reactrarrow{0pt}{0.75cm}{}{} \;  \text{\gso} \, (a,g) +  2\text{O} \, (a,g) \label{eq:c2} \;.
\end{align}

\setlength{\parindent}{6ex}
One disadvantage of using \gao\ ($s$) as the MBE source is that \gao\ does not evaporate congruently. Our thermodynamic calculations indicate that when \gao\ ($s$) is heated to a temperature where the \gso\ ($g$) that it evolves has a vapor pressure of $0.1\,\text{Pa}$ (a vapor pressure typical for MBE growth), that the \gso\ molecular-beam is only 98.0\% \gso\ molecules. The other 2\% of the beam is Ga, O$_2$, and O species.

The other disadvantage of using \gao\ ($s$) as the MBE source is that quite high effusion cell temperatures are required to evolve appreciable \fgas; temperatures in excess of $\sim 1600 \text{\utg}$,\cite{droopad2017} $\sim 1700 \text{\utg}$, \cite{passlack1998}, or $\sim 1800 \text{\utg}$ \cite{ghose2016} have been used. At such high effusion cell temperatures, crucible choices become limited and prior researchers have used iridium crucibles.\cite{passlack1998,passlack2003,ghose2016,ghose2017} \gao\ thin films synthesized utilizing an iridium crucible at an effusion cell temperature of $\sim 1700 \text{\utg}$ \cite{passlack1998} were limited to growth rates $< 0.14 \, \upmu\text{m}\,\text{hr}^{-1}$ (Ref.~\citenum{passlack2003}) with $\sim 5 \times 10^{18}\,\text{cm}^{-3}$ iridium contamination in the grown \gao\ films.\cite{passlack1998,passlack1997} These aspects of \gao\ compound sources hamper the synthesis of semiconducting \gao\ layers at growth rates exceeding $1 \, \upmu\text{m}\,\text{hr}^{-1}$ with device-relevant material properties. For comparison, the $\text{Ga} + \text{\gao}$ mixture that we describe next and have used to grow \gao\ films at growth rates exceeding $1 \, \upmu\text{m}\,\text{hr}^{-1}$ provides a \gso\ molecular-beam that is 99.98\% pure according to our thermodynamic calculations. This is for the same \gso\ vapor pressure of $0.1\,\text{Pa}$, which happens at a source temperature about $600\text{\utg}$ lower for this $\text{Ga} + \text{\gao}$ mixture than for pure \gao, enabling us to use crucibles that do not result in iridium-contaminated films.

\setlength{\parindent}{6ex}
Years ago as well as more recently, $\text{Ga} + \text{\gao}$-mixed sources producing a \gso\ molecular-beam have been studied \cite{frosch1962,hoffmann2020} and suggested as efficient suboxide sources for oxide MBE.\cite{hoffmann2020,adkison2020} Using this mixed source, a \gso\ ($g$) molecular-beam is produced by the chemical reaction
\begin{equation}
\label{eq:etch}
4\text{Ga}\,(l)  + \text{\gao}\,(s) \; \reactrarrow{0pt}{0.75cm}{}{} \;  3\text{\gso} \, (s,g) \;,
\end{equation}
with the liquid phase denoted as $l$. $S$-MBE uses the thermodynamic \cite{frosch1962} and kinetic \cite{vogt2015a} properties of $\text{Ga} + \text{\gao}$ mixtures favoring reaction (\ref{eq:etch}) under MBE conditions.

\setlength{\parindent}{6ex}
For the $S$-MBE of \gao, we explored Ga-rich and \gao-rich mixtures of $\text{Ga} + \text{\gao}$ with stoichiometries
\begin{equation}
\label{eq:garich}
5\text{Ga}\,(l)  + \text{\gao}\,(s)  \; \reactrarrow{0pt}{1.35cm}{$\kappa_{\text{Ga-rich}}$}{} \; 3\,\text{Ga$_2$O} \, (g) + \text{Ga} \, (l) \;,
\end{equation}
and
\begin{equation}
\label{eq:gaorich}
\frac{5}{2}\,\text{Ga}\,(l)  + \text{\gao}\,(s)  \; \reactrarrow{0pt}{1.35cm}{$\kappa_{\text{\gao-rich}}$}{}\; \frac{15}{8}\,\text{Ga$_2$O} \, (g) + \frac{3}{8}\,\text{\gao} \, (s) \;,
\end{equation}
respectively. The latter mixture has an oxygen mole fraction of $x(\text{O}) = 0.4$ and the properties of this \gao-rich mixture are described below. The corresponding reaction rate constants $\kappa_{\text{Ga-rich}}$ and $\kappa_{\text{\gao-rich}}$ define the production rate of \gso\ ($g$) at a given temperature $T_{\text{mix}}$ of the $\text{Ga} + \text{\gao}$ mixture. 

\setlength{\parindent}{6ex}
The flux of \gso\ ($g$) in the molecular-beam emanating from the mixed Ga + \gao\ sources is significantly larger than that of Ga ($g$) \cite{frosch1962,alcock1984} emanating from the same source. This is also true under MBE conditions.\cite{hoffmann2020,adkison2020} The resulting high ratio of $\text{\gso}/\text{Ga} \gg 1$ provides a more controllable and cleaner growth environment than accessible by decomposing a stoichiometric \gao\ source, which produces molecular-beam ratios of $\text{\gso}/\text{Ga}$, $\text{\gso}/\text{O}_2$, and $\text{\gso}/\text{O}$. Hence, the growth surface of the substrate during film growth using $S$-MBE is exposed to controllable and independently supplied molecular-beams of \gso\ and reactive O adsorbates. 

\setlength{\parindent}{6ex}
We have experienced that a \gao-rich mixture enables higher $T_{\text{mix}}$ and higher, stable \gso\ ($g$) molecular-beams than a Ga-rich mixture. This enables $S$-MBE to achieve higher \gr s. This experimental observation is confirmed by our thermodynamic calculations of the phase diagram of $\text{Ga} \, (l) + \text{\gao}\, (s)$ mixtures, which we describe next.
%%%%%%%%%%%%%%%%%%%%%%%%%%%%%%%%%%%%%%6%%%%%%%%%%%%%%%%%%%%%%%%%%%%%%%%%%.
\begin{figure}[t]
\includegraphics[width=0.475\textwidth]{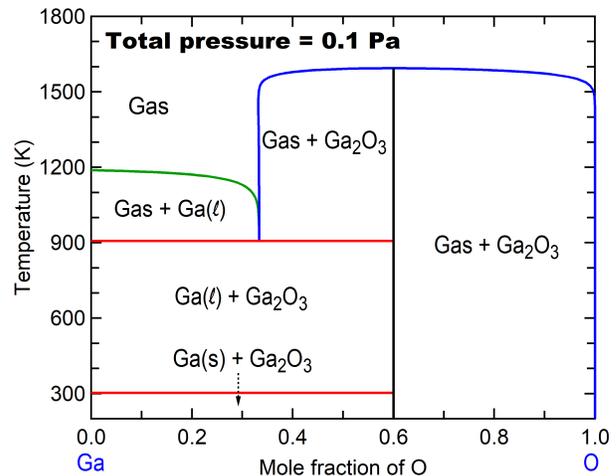}
\caption{Ga-O temperature-composition phase diagram under constant pressure $P = 0.1 \, \text{Pa}$. This phase diagram has been calculated at higher pressures by Ref.~\citenum{zinke2004}.}
\label{fig:pd1}
\end{figure}
%%%%%%%%%%%%%%%%%%%%%%%%%%%%%%%%%%%%%%%%%%%%%%%%%%%%%%%%%%%%%%%%%%%%%%%%
%%%%%%%%%%%%%%%%%%%%%%%%%%%%%%%%%%%%%%%%%%%%%%%%%%%%%%%%%%%%%%%%%%%%%%%%
\begin{figure*}[t]
\includegraphics[width=0.7\textwidth]{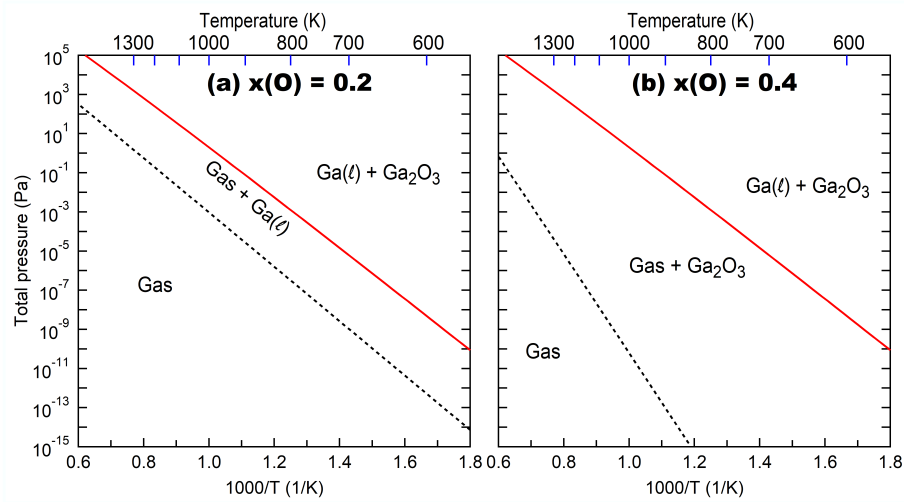}
		\caption{Ga--O pressure versus temperature ($P-T$) phase diagrams at fixed mole fractions of oxygen of $x(\text{O}) = 0.2$ [panel (a)] and $x(\text{O}) = 0.4$ [panel (b)]. These oxygen mole fractions are chosen to illustrate the difference between (a) Ga-rich mixtures and (b) \gao-rich mixtures.} 
\label{fig:tc1}
\end{figure*}
%%%%%%%%%%%%%%%%%%%%%%%%%%%%%%%%%%%%%%%%%%%%%%%%%%%%%%%%%%%%%%%%%%%%%%%%
%%%%%%%%%%%%%%%%%%%%%%%%%%%%%%%%%%%%%%%%%%%%%%%%%%%%%%%%%%%%%%%%%%%%%%%%
\begin{figure*}[t]
\includegraphics[width=0.45\textwidth]{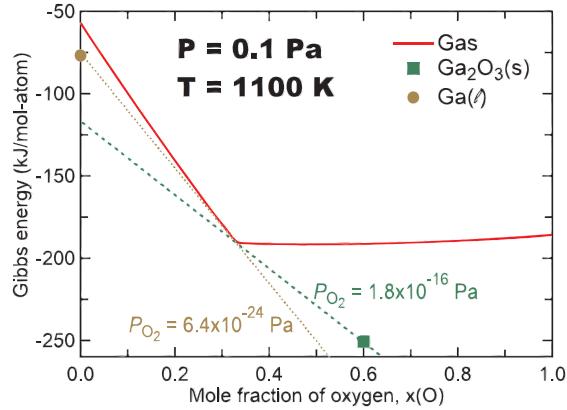}
		\caption{Gibbs energies of the gas, Ga($l$), \gao($s$) phases at temperature $T = 1100\,\text{K}$ and total pressure $P = 0.1\,\text{Pa}$. The brown dotted line shows the activity (or partial pressure) of oxygen when $0 < x(\text{O}) < 0.33$. In this range the gas phase is in equilibrium with Ga($l$) and the activity of oxygen is $6.4 \times 10^{-24}\,\text{Pa}$. The green dashed line corresponds to the case where $0.33 < x(\text{O}) < 0.6$. In this range the gas phase is in equilibrium with \gao($s$) and the activity of oxygen is $P_{\text{O}_2} = 1.8 \times 10^{-16}\,\text{Pa}$. This difference in the partial pressure of O$_2$ between the two regimes is huge and shows the advantage of growing \gao\ films from \gao-rich ($\text{Ga} + \text{\gao}$) mixtures.} 
\label{fig:tc11}
\end{figure*}
%%%%%%%%%%%%%%%%%%%%%%%%%%%%%%%%%%%%%%%%%%%%%%%%%%%%%%%%%%%%%%%%%%%%%%%%
%%%%%%%%%%%%%%%%%%%%%%%%%%%%%%%%%%%%%%%%%%%%%%%%%%%%%%%%%%%%%%%%%%%%%%%%
\begin{figure*}[t]
\includegraphics[width=0.715\textwidth]{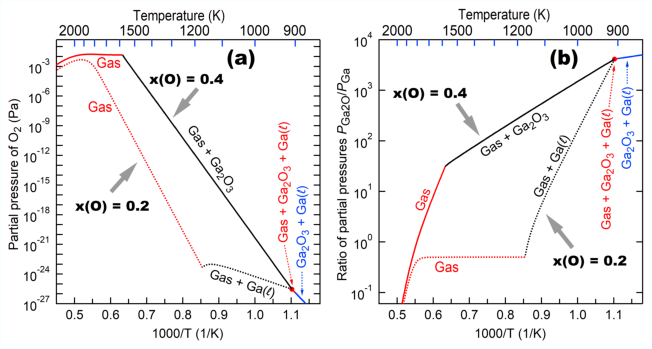}
		\caption{(a) Partial pressure of oxygen and (b) ratio of the partial pressure of \gso\ to that of Ga plotted as a function of temperature with the total pressure being $0.1\,\text{Pa}$ for the mole fractions of oxygen at $x(\text{O}) = 0.2$ (dotted lines) and $x(\text{O}) = 0.4$ (solid lines), respectively. These oxygen mole fractions are chosen to illustrate the difference between Ga-rich mixtures [$x(\text{O}) = 0.2$] and \gao-rich mixtures [$x(\text{O}) = 0.4$].} 
\label{fig:tc2}
\end{figure*}
%%%%%%%%%%%%%%%%%%%%%%%%%%%%%%%%%%%%%%%%%%%%%%%%%%%%%%%%%%%%%%%%%%%%%%%%

\setlength{\parindent}{6ex}
The calculated Ga--O phase diagram in Fig.~\ref{fig:pd1} shows that at $T_{\text{mix}}$ below the three-phase equilibrium of $\text{gas} + \text{Ga} \, (l) + \text{\gao} \, (s)$ around $907 \, \text{K}$, a two-phase region of $\text{Ga} \, (l) + \text{\gao} \, (s)$ forms, which does not change with respect to temperature or oxygen mole fraction between 0 and 0.6. Note that all thermodynamic calculations in the present work were performed using the Scientific Group Thermodata Europe (SGTE) substance database (SSUB5) \cite{sgte} within the Thermo-Calc software.\cite{anders2002} For $T_{\text{mix}} > 907 \, \text{K}$, the two-phase regions are $\text{gas} + \text{Ga} \, (l)$ when the mole fraction of oxygen is below 1/3, corresponding to what we refer to as Ga-rich mixtures, and $\text{gas} + \text{\gao} \, (s)$ when the mole fraction of oxygen is between 1/3 and 0.6, which we refer to as \gao-rich mixtures. These two-phase regions become a single gas-phase region at $T_{\text{mix}}$ of $(907 - 1189) \, \text{K}$ for Ga-rich mixtures and at $(907 - 1594) \, \text{K}$ for \gao-rich mixtures, respectively. All of these phase transition temperatures decrease with decreasing pressure \cite{zinke2004} as shown on the pressure versus temperature ($P-T$) phase diagrams in Fig.~\ref{fig:tc1}.

\setlength{\parindent}{6ex}
To contrast the difference between Ga-rich versus \gao-rich mixtures we have performed additional thermodynamic calculations at oxygen mole fractions of $x(\text{O}) = 0.2$ and $x(\text{O}) = 0.4$. These two chosen oxygen mole fractions correspond to Ga-rich and \gao-rich mixtures, respectively. In Figs.~\ref{fig:tc1}(a) and \ref{fig:tc1}(b) the solid (red) lines denote the three-phase equilibrium between $\text{gas} + \text{Ga} \, (l) + \text{\gao} \, (s)$; these are identical at $x(\text{O}) = 0.2$ and $x(\text{O}) = 0.4$. The dotted (black) lines denote the equilibrium between the gas and $\text{gas} + \text{Ga} \, (l)$ phase regions for $x(\text{O}) = 0.2$ and the gas and $\text{gas} + \text{\gao} \, (s)$ phase regions for $x(\text{O}) = 0.4$, i.e., their respective boiling temperature/pressure. 

\setlength{\parindent}{6ex}
Figure \ref{fig:tc11} shows Gibbs energies of the gas, Ga($l$), \gao($s$) phases at temperature $T = 1100\,\text{K}$ and total pressure $P = 0.1\,\text{Pa}$. There are seven distinct atomic and molecular species in the gas phase:~Ga, Ga$_2$, GaO, Ga$_2$O, O, O$_2$, and O$_3$. The kink in the Gibbs energy of the gas phase at $x(\text{O}) = 0.33$ corresponds to the composition of the Ga$_2$O species because it is the major species in the gas phase. It can be seen that the values of the oxygen activity in the $\text{gas} + \text{Ga} \, (l)$ vs.~in the $\text{gas} + \text{\gao} \, (s)$ regions differ by more than seven orders of magnitude, i.e., $6.4 \times 10^{-24}\,\text{Pa}$ vs.~$1.8 \times 10^{-16}\,\text{Pa}$ as indicated by the brown and green common tangent lines in Fig.~\ref{fig:tc11}. 

\setlength{\parindent}{6ex}
In Fig.~\ref{fig:tc2}(a) the partial pressure of oxygen in the gas phase is plotted as a function of temperature (for a total pressure of $0.1\,\text{Pa}$) for a Ga-rich mixture at $x(\text{O}) = 0.2$ and a \gao-rich mixture at $x(\text{O}) = 0.4$. It can be seen that the oxygen partial pressure in the \gao-rich mixture at $x(\text{O}) = 0.4$ is orders of magnitude higher than that at $x(\text{O}) = 0.2$ at relevant MBE growth temperatures. For example, the value of the partial pressures of oxygen at $T_{\text{mix}} = 1000\,\text{K}$ at $x(\text{O}) = 0.2$ is $5.6 \times 10^{-25}\,\text{Pa}$ and at $x(\text{O}) = 0.4$ is $4.5 \times 10^{-21}\,\text{Pa}$. The higher oxygen activity of \gao-rich mixtures compared with Ga-rich mixtures makes it easier to form fully oxidized \gao\ thin films. At lower total pressure, all lines shift to lower temperatures.

\setlength{\parindent}{6ex}
Further, our thermodynamic calculations plotted in Fig.~\ref{fig:tc2}(b) show the ratio of the partial pressures of \gso\ to Ga in the gas phase as a function of the temperature of a Ga-rich mixture [$x(\text{O}) = 0.2$] and of a \gao-rich mixture [$x(\text{O}) = 0.4$], where the total pressure is fixed at $0.1\,\text{Pa}$. The ratio of the partial pressures of \gso\ to Ga in a Ga-rich mixture with $x(\text{O}) = 0.2$ is much lower than this ratio in a \gao-rich mixture with $x(\text{O}) = 0.4$. For example, the $P_{\text{\gso}} / P_{\text{Ga}}$ ratio is 158 in a Ga-rich mixture [$x(\text{O}) = 0.2$] and 1496 in a \gao-rich mixture [$x(\text{O}) = 0.4$] at $T_{\text{mix}} = 1000\,\text{K}$. The higher \gso/Ga ratios at higher $T_{\text{mix}}$ are another reason why \gao-rich mixtures are preferred. Higher \gso/Ga ratios and the higher purity of the \gso\ molecular-beam [99.98\% \gso\ according to our calculations at $x(\text{O}) = 0.4$] mean that the \gao\ films are formed by a single-step reaction [reaction (\ref{eq:sec})] and that reaction (\ref{eq:first}) is bypassed.

\setlength{\parindent}{6ex}
We used Ga metal (7N purity) and \gao\ powder (5N purity) for the Ga + \gao\ mixtures, loaded them into a $40\,\text{cm}^{3}$ \alo\ crucible and inserted it into a commercial dual-filament, medium temperature MBE effusion cell. After mounting the effusion cell to our Veeco GEN10 MBE system and evacuating the source, we heated it up, outgased the mixture, and set our desired \gso\ flux for the growth of \gao. We measured the flux of the \gso\ ($g$) molecular-beam reaching the growth surface prior to and after growth using a quartz crystal microbalance. The film surface was monitored during growth by reflection high-energy electron diffraction (RHEED) using $13\,\text{keV}$ electrons. After growth x-ray reflectivity (XRR), optical reflectivity in a microscope (ORM),\cite{fil} scanning electron microscopy (SEM), scanning transmission electron microscopy (STEM), and secondary-ion mass spectrometry (SIMS) were used to accurately measure the thicknesses of homoepitaxial (ORM, SEM, SIMS, SEM) and heteroepitaxial (XRR, ORM, SEM, STEM, SIMS) grown \gao\ films to determine the growth rate. X-ray diffraction was performed using a four-circle x-ray diffractometer with Cu K$\alpha_1$ radiation.

\section{Results for \gao\ using $S$-MBE}
\subsection{Growth rates and growth model}
\begin{figure}[t]
\includegraphics[scale=1]{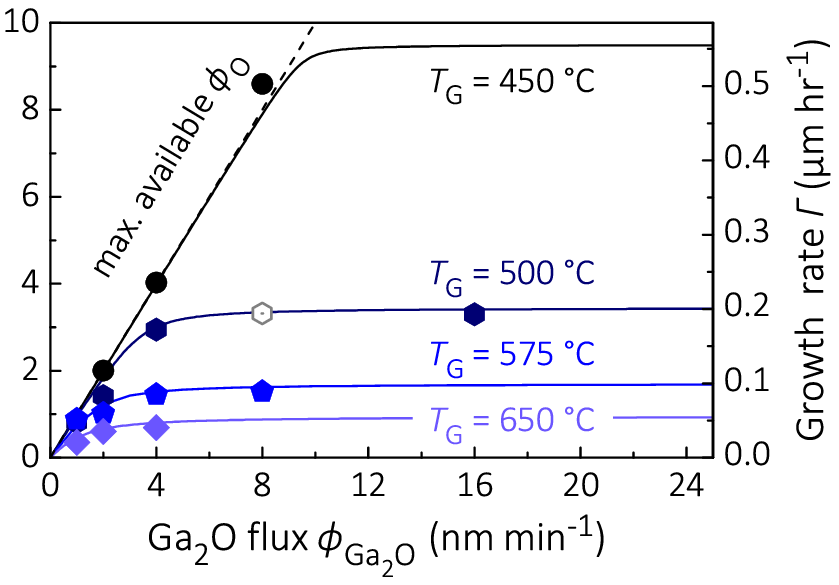}
\caption{Measured \gr\ of \gao($\bar{2}$01)/\alo(0001) as a function of \fgas\ at different \tg\ (as indicated in the figure). Solid lines are fits of our model, Eqs.~(\ref{eq:m1})--(\ref{eq:m3}), to the data. A flux of \fo\ was provided by an oxidant---a mixture of $\text{O}_2$ and approximately $80\,\%$ O$_3$ \cite{theis1997}---supplied continuously during growth at a background pressure of $1 \times 10^{-6}\,\text{Torr}$. The dashed line reveals the transition between O-rich and \gso-rich growth regimes and indicates the maximum available O flux (which equals the growth rate value of the plateau) for \gso\ to \gao\ conversion at a given \tg.}
\label{fig:gr1}
\end{figure}

\setlength{\parindent}{6ex}
Figure \ref{fig:gr1} plots the \gr\ of \gao\ as a function of \fgas\ at different \tg\ and constant \fo. The \gr s obtained follow the anticipated growth kinetics depicted in Fig.~\ref{fig:schem}(c). In the adsorption-controlled regime, an increase in \fgas\ (at otherwise constant growth parameters) does \textit{not} lead to a decrease in the \gr\ as observed for `conventional' \gao\ MBE [Fig.~\ref{fig:schem}(b)],\cite{tsai2010,vogt2016a} but instead results in a constant growth rate:~a \gr-plateau. The data clearly show that we have overcome the \gr-limiting step by using a \gso\ ($g$) suboxide molecular-beam while reducing the complexity of the \gao\ reaction kinetics from a two-step [Eqs.~(\ref{eq:first}) and (\ref{eq:sec})] to a single-step [Eq.~(\ref{eq:sec})] reaction mechanism.

\setlength{\parindent}{6ex}
The reaction kinetics of $S$-MBE for the growth of \gao\ ($s$) can be described in a similar way as `conventional' III-V [e.g., reaction (\ref{eq:gan})] and II-VI MBE. We therefore set up a simple reaction-rate model describing the growth of \gao\ ($s$) by $S$-MBE (this same model applies to other III-VI and IV-VI compounds, as well):
\begin{align}
\frac{\text{d}n_{\text{\gso}}}{\text{d}t} &= \phi_{\text{\gso}} - \kappa_{\text{\gso}} \, n_{\text{\gso}} \, n_{\text{O}}^2 - \gamma_{\text{\gso}} \, n_{\text{\gso}} \,,\label{eq:m1}
\\
\frac{\text{d}n_{\text{O}}}{\text{d}t} &= \sigma \text{\fo} - 2 \, \kappa_{\text{\gso}} \, n_{\text{\gso}} \, n_{\text{O}}^2 - \gamma_{\text{O}} \, n_{\text{O}} \,,\label{eq:m2}
\\
\frac{\text{d}n_{\text{Ga$_2$O$_3$}}}{\text{d}t} &= \Gamma =  \kappa_{\text{\gso}} \, n_{\text{\gso}} \, n_{\text{O}}^2 \label{eq:m3} \;.
\end{align}
The \gao, \gso, and O adsorbate densities are denoted as $n_{\text{\gao}}$, $n_{\text{\gso}}$, and $n_{\text{O}}$, respectively. Their time derivative is described by the operator $\text{d}/\text{d}t$. The reaction rate constant $\kappa_{\text{\gso}}$ kinetically describes the \gr\ $\Gamma$ of \gao\ ($s$) on the growth surface. The desorption rate constants of \gso\ and O adsorbates are denoted as $\gamma_{\text{\gso}}$ and $\gamma_{\text{O}}$, respectively.
%%%%%%%%%%%%%%%%%%%%%%%%%%%%%%%%%%%%%%%%%%%%%%%%%%%%%%%%%%%%%%%%%%%%%%%%.
%%%%%%%%%%%%%%%%%%%%%%%%%%%%%%%%%%%%%%%%%%%%%%%%%%%%%%%%%%%%%%%%%%%%%%%%

\setlength{\parindent}{6ex}
The flux of available O adsorbates, for \gso\ to \gao\ oxidation at a given \tg, is determined by its sticking coefficient $\sigma$ on the \gao\ growth surface and is described by a sigmoid function
\begin{equation}
\label{eq:oflux}
\sigma\bigl(\text{\tg}\bigr) = \Biggl[\sigma_0 \, \text{exp}\biggl(-\frac{\Delta \sigma}{k_{\text{B}} \bigl(\text{\tg} - \text{d}\text{\tg}\bigr)}\biggr)+1\Biggr]^{-1} \;,
\end{equation}
with dimensionless pre-factor $\sigma_0$, energy $\Delta \sigma$, and temperature off-set $\text{d}\text{\tg}$. Equation (\ref{eq:oflux}) reflects the decreasing probability of O species to adsorb as \tg\ is increased. This leads to an effectively lower surface density of active O for \gso\ oxidation and thus to lower \gr s.
%%%%%%%%%%%%%%%%%%%%%%%%%%%%%%%%%%%%%%%%%%%%%%%%%%%%%%%%%%%%%%%%%%%%%%%%.
\begin{figure}[t]
\includegraphics[scale=0.95]{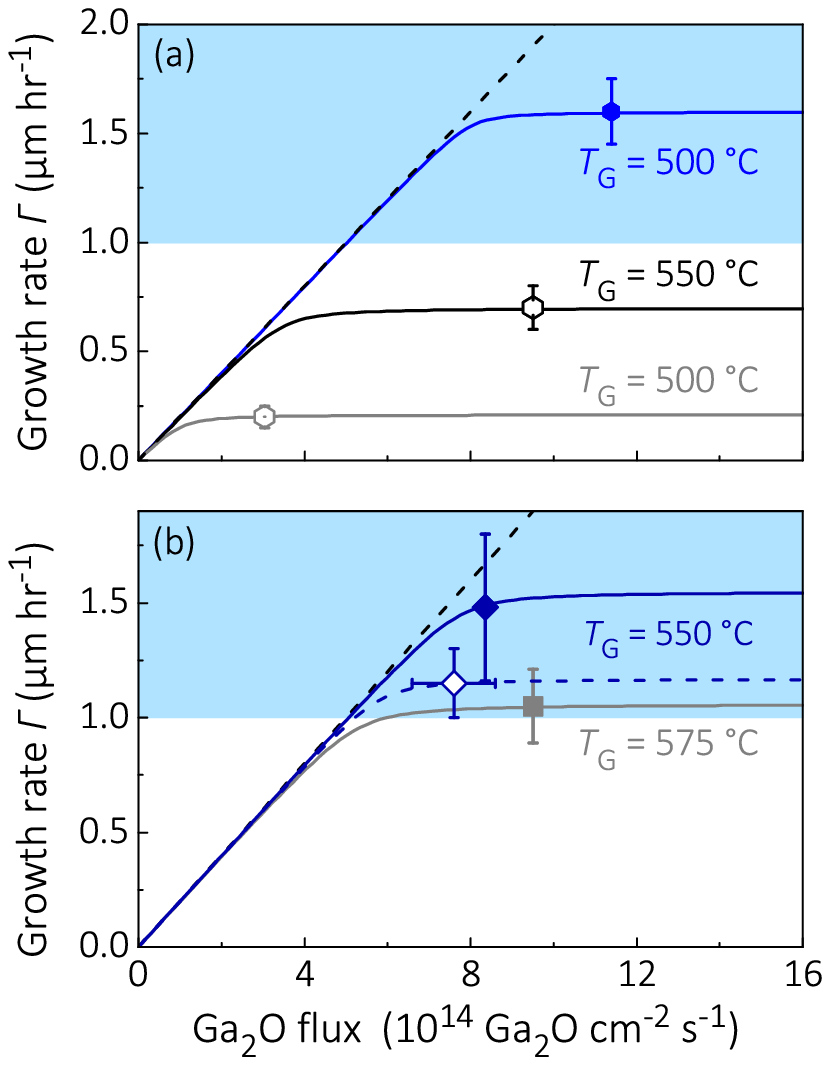}
\caption{(a) Examples of measured \gr s of $1.6\,\upmu\text{m}\,\text{hr}^{-1}$ (solid hexagon), $0.7\,\upmu\text{m}\,\text{hr}^{-1}$ (open hexagon), and $0.2\,\upmu\text{m}\,\text{hr}^{-1}$ (open-dotted hexagon; the same data point is shown in Fig.~\ref{fig:gr1} in $\text{nm}\,\text{min}^{-1}$) of \gao($\bar{2}$01) grown on \alo(0001) at \fgas\ of $11.4$, $9.5$, and $3.0 \times 10^{14}\,\text{\gso}\,\text{molecules}\,\text{cm}^{-2}\,\text{s}^{-1}$, respectively. The oxygen flux was provided by an oxidant (O$_2$ + $80\,\%$ O$_3$) background pressure of $5 \times 10^{-6}\,\text{Torr}$ (solid hexagon and open hexagon) as well as $1 \times 10^{-6}\,\text{Torr}$ (open-dotted hexagon). (b) Examples of measured \gr s of $1.5\,\upmu\text{m}\,\text{hr}^{-1}$ (solid diamond), $1.17\,\upmu\text{m}\,\text{hr}^{-1}$ (open diamond), and $1.05\,\upmu\text{m}\,\text{hr}^{-1}$ (solid square) of \gao(010) grown on \gao(010) at \fgas\ of $8.4$, $7.6$, and $9.5\times 10^{14}\,\text{\gso}\,\text{molecules}\,\text{cm}^{-2}\,\text{s}^{-1}$, respectively. The oxygen flux was provided by an oxidant (O$_2$ + $80\,\%$ O$_3$) background pressure of $5 \times 10^{-6}\,\text{Torr}$. Growth temperatures, \tg, are indicated in the figure. Lines are estimations from our model, Eqs.~(\ref{eq:m1})--(\ref{eq:m3}), including all kinetic parameters \cite{vogt2020b}. The dashed line shows the estimated intersection between the O-rich to the \gso-rich growth regime \cite{vogt2020b}. The blue shaded area indicates the adsorption-controlled \gr-regime only accessible by $S$-MBE with $\text{\gr s} \geq 1\,\upmu\text{m}\,\text{hr}^{-1}$.}
\label{fig:gr2}
\end{figure}
%%%%%%%%%%%%%%%%%%%%%%%%%%%%%%%%%%%%%%%%%%%%%%%%%%%%%%%%%%%%%%%%%%%%%%%%

\setlength{\parindent}{6ex}
For a supplied flux of \fo\ corresponding to a background pressure of $1 \times 10^{-6}\,\text{Torr}$ (involving mixtures of $\text{O}_2$ and approximately $10\,\%$ O$_3$ as well as $80\,\%$ O$_3$)\cite{theis1997}, the values of the variables given in Eq.~(\ref{eq:oflux}) are:~$\sigma_0 = 40$, $\Delta \sigma = 29 \, \text{meV}$, and $\text{d}\text{\tg} = 675\text{\utg}$. In this work, we introduce this model for $S$-MBE to demonstrate its practical value. A physical description of this model including all model parameters is given in Ref.~\citenum{vogt2020b}.
The given values are extracted by fitting the maximum \gr\ (defined as the plateau-regime) as a function of \tg, e.g., as plotted in Fig.~\ref{fig:gr1}. We find that $\sigma$ does not depend on the concentration of active O; it only depends on the partial pressure of active O. Thus, the active O may be be scaled up or down by either changing the concentration of O$_3$ in the O$_3$ beam or by changing the partial pressure of O$_3$ in the chamber. Note that O$_3$ supplies O to the surface of the growing film when it decomposes by the reaction:~O$_3$ ($g$) $\rightarrow$ O$_2$ ($g$) + O ($g$). A similar behavior of an increasing desorption or recombination rate of active O species with increasing \tg\ has also been observed during O plasma-assisted MBE using elemental Ga and O molecular-beams.\cite{vogt2016a,vogt2016b,vogtdiss} 

\setlength{\parindent}{6ex}
Based on this model, we scaled up \fo\ in order to achieve \gao\ ($s$) \gr s that exceed $1\,\upmu\text{m}\,\text{hr}^{-1}$. Figure \ref{fig:gr2}(a) demonstrates our fastest (to date) $\text{\gr}$ of $1.6\,\upmu\text{m}\,\text{hr}^{-1}$ of a $\upbeta$-\gao\ thin film grown on \alo(0001), at $\text{\tg} = 500\text{\utg}$.
For comparison, the data point plotted as an open-dotted hexagon (see also Fig.~\ref{fig:gr1}) shows the highest possible \gr\ at a five times lower active \fo\ and the same \tg. This result shows quite clearly the accuracy of our model and demonstrates the $S$-MBE of \gao\ thin films at \gr s exceeding $1 \,\upmu\text{m}\,\text{hr}^{-1}$. In addition, the growth rate values plotted in Fig.~\ref{fig:gr2}(b) were obtained by homoepitaxial growth of $\upbeta$-\gao(010) on $\upbeta$-\gao(010). The \gr\ of \gao\ on \gao(010) is 2.1 times larger than the \gr\ on \alo(0001) at similar growth conditions---e.g., as plotted in Figs.~\ref{fig:gr2}(a) [open hexagon] and \ref{fig:gr2}(b) [solid diamond], respectively. This result suggests that the growth rate of $S$-MBE grown \gao(010) and other surfaces of \gao\ may vastly exceed $1 \,\upmu\text{m}\,\text{hr}^{-1}$ in the adsorption-controlled regime. The higher growth rate is likely due to the surface-dependent adhesion energies between of \gso\ adsorbates and substrate \cite{vogtdiss,vogt2018b,mazzo2020}, similar to what has been observed for Ga adsorbates during the `conventional' MBE of \gao\ \cite{sasaki2012}. Fluctuations in \tg\ and \fgas\ for different samples and during the long duration growth of the `thick' sample (> 3 hours) are considered by the standard deviations of the measured values of \tg\ and \fgas\ as given in Fig.~\ref{fig:gr2}.
%%%%%%%%%%%%%%%%%%%%%%%%%%%%%%%%%%%%%%%%%%%%%%%%%%%%%%%%%%%%%%%%%%%%%%%%.
\begin{figure}[t]
\includegraphics[scale=0.45]{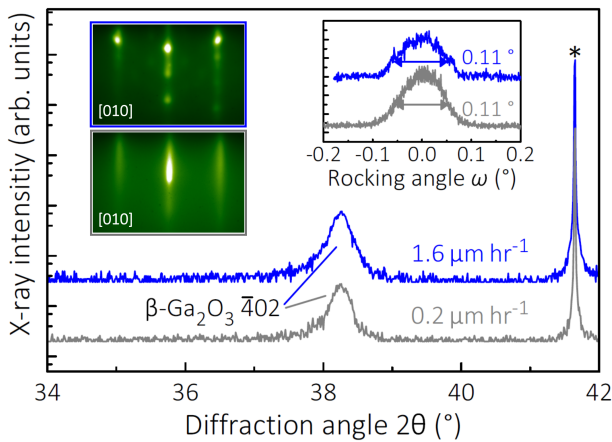}
\caption{Longitudinal XRD scans recorded for \gao\ films grown on \alo(0001) single-crystal substrates in the adsorption-controlled regime. The blue line corresponds to a film with thickness of $d = 0.15\,\upmu\text{m}$ grown at $\text{\fgas} = 11.4\times 10^{14}\,\text{\gso}\,\text{molecules}\,\text{cm}^{-2}\,\text{s}^{-1}$ where \text{\fo} was provided by an oxidant (O$_2$ + $80\,\%$ O$_3$) background pressure of $5 \times 10^{-6}\,\text{Torr}$ [see also solid blue hexagon in Fig.~(\ref{fig:gr2})(a)]. The gray line corresponds to a \gao\ film with thickness $d = 0.05\,\upmu\text{m}$ grown at $\text{\fgas} = 3.0\times 10^{14}\,\text{\gso}\,\text{molecules}\,\,\text{cm}^{-2}\,\text{s}^{-1}$ where \text{\fo} was provided by an oxidant (O$_2$ + $80\,\%$ O$_3$) background pressure of $1 \times 10^{-6}\,\text{Torr}$ [see also gray open-dotted hexagon in Fig.~(\ref{fig:gr2})(a)]. For both samples $\text{\tg}$ was $500\text{\utg}$. The reflections from the \gao\ film are identified to originate from the monoclinic $\upbeta$-phase,\cite{ahman1996} as indicated in the figure. (Inset) Transverse XRD scans across the $\bar{4}02$ peak with their FWHM indicated in the figure (same value for both films). The 0006 peaks of the \alo\ substrates are marked by an asterisk. RHEED images taken at the end of the growth along the [010] azimuth of the \gao\ films grown at \gr s of $1.6\,\upmu\text{m}\,\text{hr}^{-1}$ and $0.2\,\upmu\text{m}\,\text{hr}^{-1}$ are outlined by the blue and gray boxes, respectively.}
\label{fig:xrd}
\end{figure}
%%%%%%%%%%%%%%%%%%%%%%%%%%%%%%%%%%%%%%%%%%%%%%%%%%%%%%%%%%%%%%%%%%%%%%%% 

\subsection{Structural properties}
\setlength{\parindent}{6ex}
We investigated the impact of variable growth conditions (i.e., \fgas, \fo, and \tg) on the structural perfection of epitaxial \gao\ ($s$) films grown on \alo(0001) and \gao(010) substrates. Figure \ref{fig:xrd} shows $\theta$-$2\theta$ x-ray diffraction (XRD) scans of selected \gao\ films---the same samples depicted in Fig.~\ref{fig:gr2}(a) [solid blue hexagon and open-dotted hexagon]. The reflections of the films coincide with the $\upbeta$-\gao\ phase grown with their ($\bar{2}01$) plane parallel to the (0001) plane of the \alo\ substrate. The inset shows transverse scans (rocking curves) across the symmetric $\bar{4}02$ reflection of the same layers. The full width at half maxima (FWHM) in $\omega$ of the profiles are a measure of the out-of-plane mosaic spread of the \gao\ layer. The obtained $\Delta \omega = 0.11^{\circ} \approx 400^{\prime \prime}$ (arcseconds) does not change with \gr\ and is particularly remarkable since $\upbeta$-\gao($\bar{2}01$) films grown on \alo(0001), using elemental Ga \cite{wei2019,tsai2010} or compound \gao\ sources \cite{ghose2017}, usually show much broader line profiles in their out-of-plane crystal distributions (from $\Delta \omega \approx 0.23^{\circ}$ \cite{ghose2017} to $\Delta \omega \sim 1.00^{\circ}$)\cite{tsai2010}. Thus, the profiles in Fig.~\ref{fig:xrd} reveal a well-oriented and high quality epitaxial \gao($\bar{2}$01) thin film. Furthermore, reflection high-energy electron diffraction (RHEED) and XRR measurements reveal a sharp and well-defined interface between \gao($\bar{2}$01) and \alo\ as well as a relatively smooth surface morphology obtained by $S$-MBE. We note that in the highly adsorption-controlled regime at lower \tg\ the accumulation of \gso\ adsorbates (crystallites) on the growth surface may occur, similar to the formation of Ga droplets during GaN growth \cite{hey2000}. This effect is indicated by the slightly spotty RHEED image (outlined by the blue square) in Fig.~\ref{fig:xrd}. We have not yet optimized the growth for \gao($\bar{2}$01) films on \alo(0001) with thicknesses $\gg 1\,\upmu\text{m}$ and mapped all growth regimes (e.g., \gso\ `droplet' formation at very high \fgas) . Further investigations of the structural perfection and electrical properties of \gao\ grown by $S$-MBE need to be performed. This could be particularly interesting for the the growth of \gao\ ($s$) at even higher \gso\ ($g$) fluxes, which push even further into the adsorption-controlled regime. 
%%%%%%%%%%%%%%%%%%%%%%%%%%%%%%%%%%%%%%%%%%%%%%%%%%%%%%%%%%%%%%%%%%%%%%%%.
\begin{figure}[t]
\includegraphics[scale=0.96]{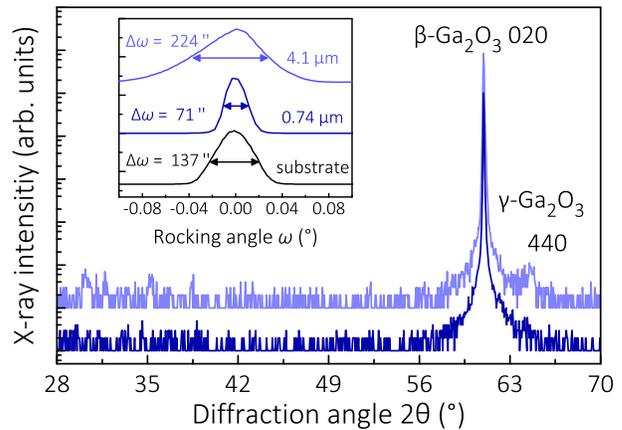}
\caption{Longitudinal XRD scans recorded for \gao\ films grown on \gao(010) single-crystal substrates in the adsorption-controlled regime. The pale blue and dark blue lines correspond to \gao\ films with thicknesses of $d = 4.1\,\upmu\text{m}$ and $d = 0.74\,\upmu\text{m}$, respectively. The reflections of the films coincide with the $\upbeta$-\gao(010) phase grown with their (010) plane parallel to the plane of the substrate. (Inset) Transverse scans across the $020$ peak of the same samples with their FWHM indicated in the figure. For comparison, a transverse scan of a single-crystalline \gao(010) substrate is also shown. The \gao(010) films (pale blue and dark blue) were grown at $\text{\fgas} = 9.1\times 10^{14}\,\text{\gso}\,\text{molecules}\,\text{cm}^{-2}\,\text{s}^{-1}$ and $\text{\tg} = 550\text{\utg}$ where \text{\fo} was provided by an oxidant (O$_2$ + $80\,\%$ O$_3$) background pressure of $5 \times 10^{-6}\,\text{Torr}$. The surface morphologies of the `thin' ($d = 0.74\,\upmu\text{m}$) and `thick' ($d = 4.1\,\upmu\text{m}$) \gao(010) films are depicted in Figs.~\ref{fig:afm}(a) and \ref{fig:afm}(b). The \gr s of the `thin' and `thick' films are depicted by the solid and open diamonds, respectively, in Fig.~\ref{fig:gr2}(b).}
\label{fig:xrdhomo}
\end{figure}
%%%%%%%%%%%%%%%%%%%%%%%%%%%%%%%%%%%%%%%%%%%%%%%%%%%%%%%%%%%%%%%%%%%%%%%% 
%%%%%%%%%%%%%%%%%%%%%%%%%%%%%%%%%%%%%%%%%%%%%%%%%%%%%%%%%%%%%%%%%%%%%%%%.
\begin{figure}[t]
\includegraphics[scale=0.825]{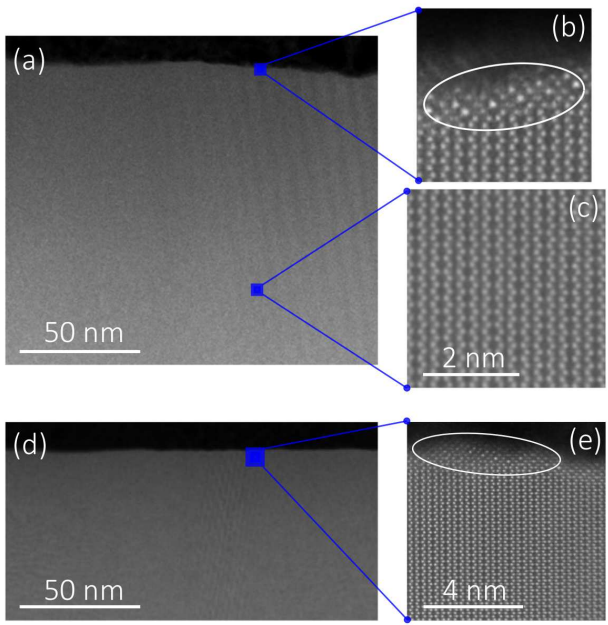}
\caption{(a)--(c) STEM images along the [001] zone axis of a \gao(010) `thin' film grown at a \gr\ of $1.05\,\upmu\text{m}\,\text{hr}^{-1}$ with thickness $d = 0.28\,\upmu\text{m}$ [this is the same sample depicted by the solid square in Fig.~\ref{fig:gr2}(b)]. The surface morphology of this same sample is shown in Fig.~\ref{fig:afm}(c). (d)--(e) STEM images of a \gao(010) `thick' film at \gr\ of $1.17\,\upmu\text{m}\,\text{hr}^{-1}$ with thickness $4.1\,\upmu\text{m}$ [this is the same sample depicted by the open diamond in Fig.~\ref{fig:gr2}(b) and pale blue line in Fig.~\ref{fig:xrdhomo}]. The surface morphology of this film is depicted in Fig.~\ref{fig:afm}(b).
No large-scale defects or dislocations are observed within either layers [panels (a) and (d)]. The \gao\ films consist only of the $\upbeta$-\gao(010) phase [panel (c) and (e)], except for a thin $\upgamma$-\gao\ phase at the top surface [highlighted by a white circle in (b) and (e)].}
\label{fig:tem}
\end{figure}
%%%%%%%%%%%%%%%%%%%%%%%%%%%%%%%%%%%%%%%%%%%%%%%%%%%%%%%%%%%%%%%%%%%%%%%% 
%%%%%%%%%%%%%%%%%%%%%%%%%%%%%%%%%%%%%%%%%%%%%%%%%%%%%%%%%%%%%%%%%%%%%%%
\begin{figure*}[t]
\includegraphics[width=0.875\textwidth]{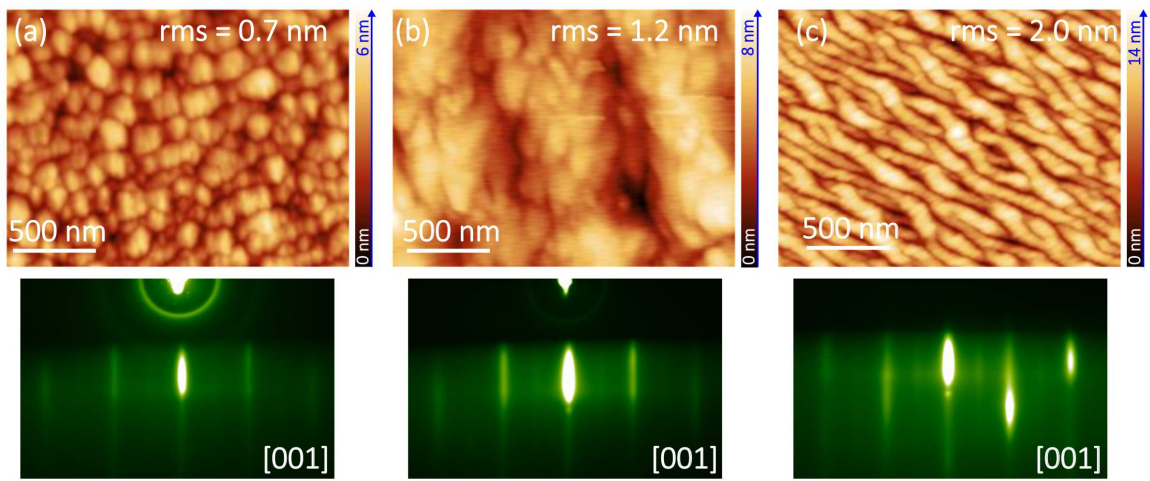}
		\caption{(a)--(c) Surface morphologies obtained by AFM for \gao(010) surfaces grown by $S$-MBE. The rms roughness of the surfaces are indicated on the figures. The XRD patterns of the same layers as shown in (a) and (b) are plotted in Fig.~\ref{fig:xrdhomo} as dark blue and pale blue lines, respectively. The \gr s of the films shown in (a), (b) and (c), are depicted in Fig.~\ref{fig:gr2} as solid diamond, open diamond, and solid square, respectively. The thicknesses of the films in (a) and (c) are $d = 0.74\,\upmu\text{m}$ and the thickness of the film with the morphology shown in (b) is $d = 4.1\,\upmu\text{m}$. \tg\ was set to $550\text{\utg}$ for the films shown in (a) and (b) and to $\text{\tg} = 575\text{\utg}$ for the film plotted in (c). RHEED images of the corresponding \gao\ film taken at the end of growth along the [001] azimuth are displayed below the respective AFM images.} 
\label{fig:afm}
\end{figure*}
%%%%%%%%%%%%%%%%%%%%%%%%%%%%%%%%%%%%%%%%%%%%%%%%%%%%%%%%%%%%%%%%%%%%%%%

\setlength{\parindent}{6ex}
We performed $S$-MBE for homoepitaxial $\upbeta$-\gao(010) films grown on $\upbeta$-\gao(010) substrates. Figure \ref{fig:xrdhomo} shows the $\theta$-$2\theta$ XRD scans of two selected \gao(010) films grown under the same growth conditions. The $\theta$-$2\theta$ XRD profiles of the \gao(010) film with thickness $d = 0.74\,\upmu\text{m}$ (plotted in dark blue) and the one of the substrate (data not shown) coincide. The \gao(010) layer with $d = 4.1\,\upmu\text{m}$ (depicted as pale blue) also shows small contributions of the meta stable $\upgamma$-\gao\ phase. The inset of Fig.~\ref{fig:xrdhomo} shows the respective rocking curves across the symmetric $020$ reflections of the same films as plotted in the main graph of Fig.~\ref{fig:xrdhomo}.
The obtained FWHM of the rocking curve of the film with $d = 0.74\,\upmu\text{m}$ is comparable to the one obtained for the bare \gao(010) substrate (depicted as a black line). [Note that the measured XRD spectra were obtained on different $10 \times 10\,\text{mm}^2$ substrates which were all cut from the same 1" diameter \gao(010) wafer from Synoptics.] The rocking curve of the `thick' film with $d = 4.1\,\upmu\text{m}$ is considerably broader than the rocking curve detected for the `thin' \gao(010) film with $d = 0.74\,\upmu\text{m}$. We attribute the different rocking curve widths measured to the non-uniformity in the crystalline perfection across the 1" diameter \gao\ substrate on which these measurements were made.

\setlength{\parindent}{6ex}
STEM of a `thin' \gao(010) film with $d = 0.28\,\upmu\text{m}$ (grown under similar conditions as the samples shown in Fig.~\ref{fig:xrdhomo}) and the `thick' film with $d = 4.1\,\upmu\text{m}$ (same sample as plotted as pale blue line in Fig.~\ref{fig:xrdhomo}) are shown in Figs.~\ref{fig:tem}(a)--\ref{fig:tem}(c) and Figs.~\ref{fig:tem}(d)--\ref{fig:tem}(e), respectively. Both samples show a clear, uniform, and single-crystalline $\upbeta$-\gao(010) film. The vertical banding in Figs.~\ref{fig:tem}(a) and \ref{fig:tem}(d) are moire fringes between the in-focus portion of the crystal lattice and the finite pixel sampling of the STEM image. Defects such as dislocations or strain fields would have distorted the fringes away from straight lines, indicating an absence of such features. Only a thin $\sim 1 \, \text{nm}$ thick $\upgamma$-\gao(110) phase at the top of the surfaces of their \gao(010)/\gao(010) structures can be seen, as marked by white circles in Figs.~\ref{fig:tem}(b) and \ref{fig:tem}(e). 

\setlength{\parindent}{6ex}
The surface morphology of \gao(010) films grown by $S$-MBE at growth rates $> 1\,\upmu\text{m}\,\text{hr}^{-1}$ were investigated by atomic force microscopy (AFM) and are plotted in Figs.~\ref{fig:afm}(a)--\ref{fig:afm}(c). The root mean square (rms) roughness of the `thin' film with $d = 0.74\,\upmu\text{m}$ is lower than the one measured for the `thick' film with $d = 4.1\,\upmu\text{m}$. This evolution in rms roughness follows the same trend as observed by XRD scans of the same layers (dark blue and pale blue lines in the inset of Fig.~\ref{fig:xrdhomo}), i.e., a slight decrease in crystal quality with increasing film thickness of the \gao(010)/\gao(010) structures. 
%%%%%%%%%%%%%%%%%%%%%%%%%%%%%%%%%%%%%%%%%%%%%%%%%%%%%%%%%%%%%%%%%%%%%%%%.
\begin{figure}[b]
\includegraphics[scale=0.975]{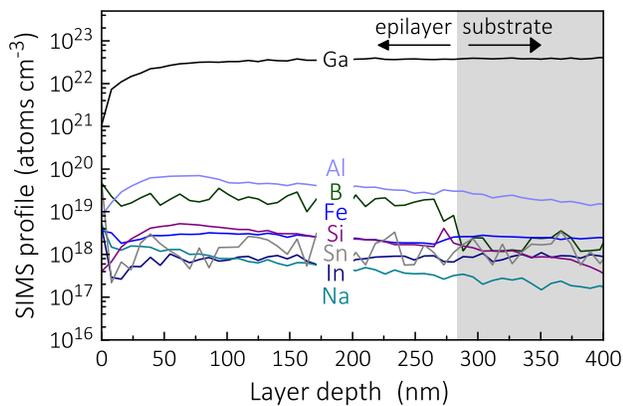}
\caption{SIMS of a \gao(010) thin film grown at $1.05\,\upmu\text{m}\,\text{hr}^{-1}$ [this is the same sample depicted by the solid square in Fig.~\ref{fig:gr2}(b)]. The atomic structure of this film and its surface morphology are shown in Figs.~\ref{fig:tem}(a)--\ref{fig:tem}(c) and \ref{fig:afm}(c), respectively. No significant impurity incorporation could be detected. Gray and white areas show the SIMS profile of the \gao(010) thin film and the Fe-doped \gao(010) substrate, respectively.}
\label{fig:sims}
\end{figure}
%%%%%%%%%%%%%%%%%%%%%%%%%%%%%%%%%%%%%%%%%%%%%%%%%%%%%%%%%%%%%%%%%%%%%%%% 

\subsection{Impurities}
\setlength{\parindent}{6ex}
We investigated the incorporation of impurities into the \gao(010) thin films grown with \gr s $> 1\, \upmu\text{m}\,\text{hr}^{-1}$ by SIMS. Figure \ref{fig:sims} shows the SIMS profile of the same film as plotted in Figs.~\ref{fig:gr2} (solid square), Fig.~\ref{fig:tem}, and Fig.~\ref{fig:afm}(c). This profile reveals that the \gao-rich ($\text{Ga} + \text{\gao}$) mixtures employed lead to \gao(010) thin films with low impurity incorporation. Only a slight increase of Al impurities with increasing film thickness and a slight incorporation of B are detected. These impurities likely originate from our use of an \alo\ crucible for the \gao-rich ($\text{Ga} + \text{\gao}$) mixture. We note that we have also used pyrolytic boron nitride (pBN) crucibles for the $\text{Ga} + \text{\gao}$ mixture, but find high concentrations of B in the grown films by SIMS ($\sim 10^{20}\,\text{B}\,\text{cm}^{-3}$) when the background pressure of a mixture of $\text{O}_2 + 80\%\text{O}_3$ is $P_{\text{O}} = 5 \times 10^{-6}\,\text{Torr}$. We attribute this to the oxidation of the surface of the pBN crucible to B$_2$O$_3$ at the high oxidant pressures used. At the $T_{\text{mix}} = 1020\,^{\circ}\text{C}$ used for growth, the vapor pressure of B$_2$O$_3$ is significant.\cite{adkison2020} The small Si peak measured at the film-substrate interface originates from unintentional incorporated Si at the substrate surface. Note, we have tried \gso-polishing (for the first time) to remove the Si from the surface prior to growth. Our observation is that \gso-polishing does not provide the same reduction in Si contamination at the sample surface as can be accomplished by Ga-polishing.\cite{ahmadi2017a}. 

\setlength{\parindent}{6ex}
Our SIMS results show that the low effusion cell temperatures and \gao-rich ($\text{Ga} + \text{\gao}$) mixtures employed for $S$-MBE---in order to produce such high \gso\ fluxes to grow \gao\ with growth rates exceeding $> 1\, \upmu\text{m}\,\text{hr}^{-1}$---do not lead to significant impurity incorporation into the grown \gao(010) films. This is an advantage of $S$-MBE compared to the growth \gao\ from a crucible containing pure \gao. Using a \gao\ compound source at extremely high effusion cell temperatures ($\sim 1700\,\text{\utg}$)\cite{passlack1998}, not only produces a flux containing a relatively low \gso\ molecular-beam resulting in low \gao\ film growth rates, but also results in films contaminated with iridium.\cite{passlack1997,passlack1998,passlack2003} Nonetheless, electrical transport properties are extremely sensitive to impurities and measurements of mobility in doped \gao\ films grown by $S$-MBE remain to be performed. It could turn out that a higher purity \gao\ powder will be needed than the 5N \gao\ powder we have used in this study. 

\subsection{Summary}
\setlength{\parindent}{6ex}
The growth rates we have achieved by $S$-MBE are more than one order of magnitude faster than what has been reported for the growth of \gao\ films from pure \gao\ sources.\cite{passlack2003}

The quality of the homoepitaxial $\upbeta$-\gao(010) films (with thickness $> 4\,\upmu\text{m}$) assessed by XRD (Fig.~\ref{fig:xrdhomo}), STEM (Fig.~\ref{fig:tem}), AFM (Fig.~\ref{fig:afm}) and SIMS (Fig.~\ref{fig:sims}), reveal that $S$-MBE with \gr s $> 1\, \upmu\text{m}\,\text{hr}^{-1}$ is competitive to other industrial relevant synthesis methods [such as (MO)CVD] for the growth of vertical \gao-based structures with thicknesses in the $\upmu\text{m}$-range. 

\setlength{\parindent}{6ex}
Based on our model and experimental results, we anticipate growth rates up to $5 \, \upmu\text{m}\,\text{hr}^{-1}$ on \gao(010) and other growth surfaces to be possible by $S$-MBE. This estimation is based on the physical MBE limit:~the mean free path $\lambda$ of the species (e.g., \gso\ and O$_3$) emanating from their sources to the target. In our estimate we have used an upper limit for the O partial pressure of $P_{\text{O}} \sim 2 \times 10^{-4} \, \text{Torr}$ [resulting in $\lambda \sim 0.1\,\text{m}$] \cite{schlom1995} and a lower \tg\ limit of $\text{\tg} \geq 725\,\text{\utg}$ [required for the adsorbed species (e.g., \gso\ and O) to crystallize into a homoepitaxial film of \gao].

\section{Outlook and alternatives of $S$-MBE}
\setlength{\parindent}{6ex}
We have demonstrated the growth of high quality \gao\ ($s$) thin films by $S$-MBE in the adsorption-controlled regime using $\text{Ga} \, (l) + \text{\gao} \, (s)$ mixtures. The high $\text{\gr} \gg 1 \, \upmu\text{m}\,\text{hr}^{-1}$, and unparalleled crystal quality of the homoepitaxial and heteroepitaxial structures obtained (with $d \gg 1\, \upmu\text{m}$) suggest the possibility of unprecedented mobilities of \gao\ thin films containing $n$-type donors (Sn, Ge, Si) grown by $S$-MBE. 

\setlength{\parindent}{6ex}
We have also developed Sn + \sno\ and Ge + GeO$_2$ mixtures in order to produce SnO ($g$) and GeO ($g$) beams for use as $n$-type donors in \gao-based heterostructures. Furthermore, we have grown \sno\ using a Sn + \sno\ mixture.\cite{hoffmann2020} Moreover, we have grown \gao\ doped with SnO using \gso\ and SnO beams and achieved controllable Sn-doping levels in these \gao\ films.\cite{vogt2020a} Nevertheless, the improvement of the $n$-type mobilities obtained during $S$-MBE, at $\text{\gr s} > 1 \, \upmu\text{m}\,\text{hr}^{-1}$, still needs to be demonstrated and shown to exceed the state-of-the-art mobilities in \gao\ films grown by `conventional' MBE.\cite{ahmadi2017}

\setlength{\parindent}{6ex}
Our comprehensive thermodynamic analysis of the volatility of 128 binary oxides plus additional two-phase mixtures of metals with their binary oxides,\cite{adkison2020} e.g., Ga + \gao, have led us to recognize additional systems appropriate for growth by $S$-MBE.  This thermodynamic knowledge coupled with our understanding of the $S$-MBE growth of \gao\ enabled us to develop In + \ino\ and Ta + Ta$_2$O$_5$ mixtures from which we have grown high-quality bixbyite \ino\ \cite{vogt2020b, hensling2020} and \ino:\sno\ (ITO, with up to $30\%$ Sn) \cite{vogt2020b, hensling2020} as well as rutile TaO$_2$ \cite{barone2020} by $S$-MBE, respectively.

\setlength{\parindent}{6ex}
Growing thin films with very high crystalline qualities at $\text{\gr s} > 1\,\upmu\text{m}\,\text{hr}^{-1}$ by using suboxide molecular-beams---with up to $5\,\upmu\text{m}\,\text{hr}^{-1}$ anticipated \gr s by our model---will make MBE competitive to other established synthesis methods, such as CVD \cite{rafique2018} or MOVPE.\cite{zhang2019} The \tg\ that we have demonstrated for high quality \gao\ layers grown by $S$-MBE is significantly lower than what has been demonstrated for the growth of high quality \gao\ films by CVD or MOVPE. This makes $S$-MBE advantageous for BEOL processing. Additionally, \gao\ grown with a vast excess of \gso\ ($g$) and high oxygen activity in \gao-rich mixtures may suppress Ga vacancies in the \gao\ layers formed, which are believed to act a compensating acceptors \cite{zacherle2013,korhonen2015}---potentially improving the electrical performance of $n$-type \gao-based devices significantly. 

\setlength{\parindent}{6ex}
The development of Al + \alo\ mixtures for the growth of epitaxial \alo\ and \algao\ at comparably high \gr s by $S$-MBE is foreseeable. In order to fabricate vertical high-power devices, thin film thicknesses in the micrometer range are desired. $S$-MBE allows the epitaxy of such devices in relatively short growth times (i.e., within a few hours as demonstrated for \gao(010) in this work) while maintaining nanometer scale smoothness. In addition, the use of a Al$_2$O ($g$) and \gso\ ($g$) molecular-beams during \algao\ $S$-MBE may also extend its growth domain towards higher adsorption-controlled regimes---being beneficial for the performance of \algao-based heterostructure devices.

\setlength{\parindent}{6ex}
Our demonstration of high quality films of \gao, \gao\ doped with SnO,\cite{vogt2020a} \ino,\cite{vogt2020b,hensling2020} ITO,\cite{vogt2020b,hensling2020} TaO$_2$,\cite{barone2020} LaInO$_3$,\cite{park2020} and LaAlO$_3$,\cite{eren2020} suggests that this synthesis-science approach---utilizing a combination of thermodynamics to identify which suboxides can be produced in molecular-beams in combination with a kinetic model of the growth process---can be applied to a wide-range of oxide compounds.\cite{adkison2020} We anticipate $S$-MBE to be applicable to all materials that form via intermediate reaction products (a~\textit{subcompound}). Examples following this reasoning include ZrO$_2$, Pb(Zr,Ti)O$_3$, and (Hf,Zr)O$_2$ all via the supply of a molecular-beam of ZrO (predicted by our thermodynamic calculations,\cite{adkison2020}) Ga$_2$Se$_3$ via Ga$_2$Se,\cite{gamal1996,teraguchi1991,vogt2018b} In$_2$Se$_3$ through In$_2$Se,\cite{greenberg1973,okamoto1997,vogt2018b} In$_2$Te$_3$ by In$_2$Te,\cite{golding1989,vogt2018b} or Sn$_2$Se via SnSe.\cite{bhatt1989,vogt2018b}

\section{Acknowledgments}
\setlength{\parindent}{6ex}
We thank J. D. Blevins for the \gao(010) substrates from Synoptics, Inc. used in this study.
K.~A., C.~S.~C., J.~P.~M., D.~J., H.~G.~X., D.~A.~M., and D.~G.~S.~acknowledge support from the  AFOSR/AFRL ACCESS Center of Excellence under Award No. FA9550-18-1-0529. J.~P.~M. also acknowledges from National Science Foundation within a Graduate Research Fellowship under Grant No. DGE-1650441. P.~V.~acknowledges support from ASCENT, one of six centers in JUMP, a Semiconductor Research Corporation (SRC) program sponsored by DARPA. F.~V.~E.~H. acknowledges support from the Alexander von Humboldt foundation in form of a Feodor Lynen fellowship. F.V.E.H and H.~P.~acknowledge support from the National Science Foundation (NSF) [Platform for the Accelerated Realization, Analysis and Discovery of Interface Materials (PARADIM)] under Cooperative Agreement No.~DMR-1539918. J.~P.~acknowledges support from the Air Force Office of Scientific Research under Award No.~FA9550-20-1-0102. S.~L.~S. and Z.~K.~L.~acknowledge the support of NSF through Grant No.~CMMI-1825538. This work made use of the Cornell Center for Materials Research (CCMR) Shared Facilities, which are supported through the NSF MRSEC Program (No.~DMR-1719875). Substrate preparation was performed in part at the Cornell NanoScale Facility, a member of the National Nanotechnology Coordinated Infrastructure (NNCI), which is supported by the NSF (Grant No.~ECCS-1542081).

\setlength{\parindent}{6ex}
Competing interests:~The authors P.~V., D.~G.~S., F.~V.~E.~H., K.~A., Z.-K.~L., B.~J.~B., S.-L.~S., Cornell University (D-9573) and the Pennsylvania State University (2020-5155) have filed a patent entitled:~Adsorption-Controlled Growth of \gao\ by Suboxide Molecular-Beam Epitaxy.

\section{Data availability}
\setlength{\parindent}{6ex}
No data that support the findings in this study are required and provided. 
\bibliography{references}
\end{document}